
%
\def\unlockat{\catcode`\@=11}
\def\lockat{\catcode`\@=12}
\unlockat
\def\d@f@ult{} \newif\ifamsfonts \newif\ifafour
\def\m@ssage{\immediate\write16}  \m@ssage{}
\m@ssage{hep-th preprint macros.  Last modified 16/10/92 (jmf).}
\message{These macros work with AMS Fonts 2.1 (available via ftp from}
\message{e-math.ams.com).  If you have them simply hit "return"; if}
\message{you don't, type "n" now: }
\endlinechar=-1  
\read-1 to\@nswer
\endlinechar=13
\ifx\@nswer\d@f@ult\amsfontstrue
    \m@ssage{(Will load AMS fonts.)}
\else\amsfontsfalse\m@ssage{(Won't load AMS fonts.)}\fi
\message{The default papersize is A4.  If you use US 8.5" x 11"}
\message{type an "a" now, else just hit "return": }
\endlinechar=-1  
\read-1 to\@nswer
\endlinechar=13
\ifx\@nswer\d@f@ult\afourtrue
    \m@ssage{(Using A4 paper.)}
\else\afourfalse\m@ssage{(Using US 8.5" x 11".)}\fi
\nonstopmode
%
%

\font\twelverm=cmr12
\font\ninerm=cmr9
\font\sixrm=cmr6
\font\fourteenbf=cmbx12 scaled\magstep1
\font\twelvebf=cmbx12
\font\ninebf=cmbx9
\font\sixbf=cmbx6
\font\fourteeni=cmmi12 scaled\magstep1      \skewchar\fourteeni='177
\font\twelvei=cmmi12                        \skewchar\twelvei='177
\font\ninei=cmmi9                           \skewchar\ninei='177
\font\sixi=cmmi6                            \skewchar\sixi='177
\font\fourteensy=cmsy10 scaled\magstep2     \skewchar\fourteensy='60
\font\twelvesy=cmsy10 scaled\magstep1       \skewchar\twelvesy='60
\font\ninesy=cmsy9                          \skewchar\ninesy='60
\font\sixsy=cmsy6                           \skewchar\sixsy='60
\font\fourteenex=cmex10 scaled\magstep2
\font\twelveex=cmex10 scaled\magstep1

\ifamsfonts
   \font\ninex=cmex9
   
   \font\sixex=cmex7 at 6pt
   
\else
   \font\ninex=cmex10 at 9pt
   
   \font\sixex=cmex10 at 6pt
   
\fi
\font\fourteensl=cmsl10 scaled\magstep2
\font\twelvesl=cmsl10 scaled\magstep1

\font\sevensl=cmsl10 at 7pt
\font\sixsl=cmsl10 at 6pt

\font\fourteenit=cmti12 scaled\magstep1
\font\twelveit=cmti12

\font\fourteentt=cmtt12 scaled\magstep1
\font\twelvett=cmtt12
\font\fourteencp=cmcsc10 scaled\magstep2
\font\twelvecp=cmcsc10 scaled\magstep1

\ifamsfonts
   
\else
   
\fi
\newfam\cpfam
\font\fourteenss=cmss12 scaled\magstep1
\font\twelvess=cmss12
\font\tenss=cmss10
\font\niness=cmss9

\font\sevenss=cmss8 at 7pt
\font\sixss=cmss8 at 6pt
\newfam\ssfam
\newfam\msafam \newfam\msbfam \newfam\eufam
\ifamsfonts
 \font\fourteenmsa=msam10 scaled\magstep2
 \font\twelvemsa=msam10 scaled\magstep1
 \font\tenmsa=msam10
 \font\ninemsa=msam9
 \font\sevenmsa=msam7
 \font\sixmsa=msam6
 \font\fourteenmsb=msbm10 scaled\magstep2
 \font\twelvemsb=msbm10 scaled\magstep1
 \font\tenmsb=msbm10
 \font\ninemsb=msbm9
 \font\sevenmsb=msbm7
 \font\sixmsb=msbm6
 \font\fourteeneu=eufm10 scaled\magstep2
 \font\twelveeu=eufm10 scaled\magstep1
 \font\teneu=eufm10
 \font\nineeu=eufm9
 
 \font\seveneu=eufm7
 \font\sixeu=eufm6
 \def\hexnumber@#1{\ifnum#1<10 \number#1\else
  \ifnum#1=10 A\else\ifnum#1=11 B\else\ifnum#1=12 C\else
  \ifnum#1=13 D\else\ifnum#1=14 E\else\ifnum#1=15 F\fi\fi\fi\fi\fi\fi\fi}
 \def\hexmsa{\hexnumber@\msafam}
 \def\hexmsb{\hexnumber@\msbfam} 
\fi
\newdimen\b@gheight             \b@gheight=12pt
\newcount\f@ntkey               \f@ntkey=0
\def\f@m{\afterassignment\samef@nt\f@ntkey=}
\def\samef@nt{\fam=\f@ntkey \the\textfont\f@ntkey\relax}
\def\rm{\f@m0 }
\def\mit{\f@m1 }
\def\cal{\f@m2 }
\def\it{\f@m\itfam}
\def\sl{\f@m\slfam}
\def\bf{\f@m\bffam}
\def\tt{\f@m\ttfam}
\def\caps{\f@m\cpfam}
\def\ssf{\f@m\ssfam}
\ifamsfonts
 \def\msa{\f@m\msafam}
 \def\msb{\f@m\msbfam} \let\bb=\msb
 \def\eu{\f@m\eufam}
\else
 \let \bb=\bf \let\eu=\bf
\fi
\def\fourteenpoint{\relax
    \textfont0=\fourteencp          \scriptfont0=\tenrm
      \scriptscriptfont0=\sevenrm
    \textfont1=\fourteeni           \scriptfont1=\teni
      \scriptscriptfont1=\seveni
    \textfont2=\fourteensy          \scriptfont2=\tensy
      \scriptscriptfont2=\sevensy
    \textfont3=\fourteenex          \scriptfont3=\twelveex
      \scriptscriptfont3=\tenex
    \textfont\itfam=\fourteenit     \scriptfont\itfam=\tenit
    \textfont\slfam=\fourteensl     \scriptfont\slfam=\tensl
      \scriptscriptfont\slfam=\sevensl
    \textfont\bffam=\fourteenbf     \scriptfont\bffam=\tenbf
      \scriptscriptfont\bffam=\sevenbf
    \textfont\ttfam=\fourteentt
    \textfont\cpfam=\fourteencp
    \textfont\ssfam=\fourteenss     \scriptfont\ssfam=\tenss
      \scriptscriptfont\ssfam=\sevenss
    \ifamsfonts
       \textfont\msafam=\fourteenmsa     \scriptfont\msafam=\tenmsa
         \scriptscriptfont\msafam=\sevenmsa
       \textfont\msbfam=\fourteenmsb     \scriptfont\msbfam=\tenmsb
         \scriptscriptfont\msbfam=\sevenmsb
       \textfont\eufam=\fourteeneu     \scriptfont\eufam=\teneu
         \scriptscriptfont\eufam=\seveneu \fi
    \samef@nt
    \b@gheight=14pt
    \setbox\strutbox=\hbox{\vrule height 0.85\b@gheight
                                depth 0.35\b@gheight width\z@ }}
\def\twelvepoint{\relax
    \textfont0=\twelverm          \scriptfont0=\ninerm
      \scriptscriptfont0=\sixrm
    \textfont1=\twelvei           \scriptfont1=\ninei
      \scriptscriptfont1=\sixi
    \textfont2=\twelvesy           \scriptfont2=\ninesy
      \scriptscriptfont2=\sixsy
    \textfont3=\twelveex          \scriptfont3=\ninex
      \scriptscriptfont3=\sixex
    \textfont\itfam=\twelveit    
    \textfont\slfam=\twelvesl    
      \scriptscriptfont\slfam=\sixsl
    \textfont\bffam=\twelvebf     \scriptfont\bffam=\ninebf
      \scriptscriptfont\bffam=\sixbf
    \textfont\ttfam=\twelvett
    \textfont\cpfam=\twelvecp
    \textfont\ssfam=\twelvess     \scriptfont\ssfam=\niness
      \scriptscriptfont\ssfam=\sixss
    \ifamsfonts
       \textfont\msafam=\twelvemsa     \scriptfont\msafam=\ninemsa
         \scriptscriptfont\msafam=\sixmsa
       \textfont\msbfam=\twelvemsb     \scriptfont\msbfam=\ninemsb
         \scriptscriptfont\msbfam=\sixmsb
       \textfont\eufam=\twelveeu     \scriptfont\eufam=\nineeu
         \scriptscriptfont\eufam=\sixeu \fi
    \samef@nt
    \b@gheight=12pt
    \setbox\strutbox=\hbox{\vrule height 0.85\b@gheight
                                depth 0.35\b@gheight width\z@ }}
\twelvepoint
%
%
\baselineskip = 15pt plus 0.2pt minus 0.1pt 
\lineskip = 1.5pt plus 0.1pt minus 0.1pt
\lineskiplimit = 1.5pt
\parskip = 6pt plus 2pt minus 1pt
\interlinepenalty=50
\interfootnotelinepenalty=5000
\predisplaypenalty=9000
\postdisplaypenalty=500
\hfuzz=1pt
\vfuzz=0.2pt
\dimen\footins=24 truecm 
\ifafour
 \hsize=16cm \vsize=22cm
\else
 \hsize=6.5in \vsize=9in
\fi
%
%
\skip\footins=\medskipamount
\newcount\fnotenumber
\def\clearfnotenumber{\fnotenumber=0} \clearfnotenumber
\def\fnote{\global\advance\fnotenumber by1 \generatefootsymbol
 \footnote{$^{\footsymbol}$}}
\def\fd@f#1 {\xdef\footsymbol{\mathchar"#1 }}
\def\generatefootsymbol{\iffrontpage\ifcase\fnotenumber
\or \fd@f 279 \or \fd@f 27A \or \fd@f 278 \or \fd@f 27B
\else  \fd@f 13F \fi
\else\xdef\footsymbol{\the\fnotenumber}\fi}
%
%
\newcount\secnumber \newcount\appnumber
\def\clearappnumber{\appnumber=64} \def\clearsecnumber{\secnumber=0}
\clearsecnumber \clearappnumber
\newif\ifs@c 
\newif\ifs@cd 
\s@cdtrue 
\def\unsectioned{\s@cdfalse\let\section=\subsection}
\newskip\sectionskip         \sectionskip=\medskipamount
\newskip\headskip            \headskip=8pt plus 3pt minus 3pt
\newdimen\sectionminspace    \sectionminspace=10pc
\def\Titlestyle#1{\par\begingroup \interlinepenalty=9999
     \leftskip=0.02\hsize plus 0.23\hsize minus 0.02\hsize
     \rightskip=\leftskip \parfillskip=0pt
     \advance\baselineskip by 0.5\baselineskip
     \hyphenpenalty=9000 \exhyphenpenalty=9000
     \tolerance=9999 \pretolerance=9000
     \spaceskip=0.333em \xspaceskip=0.5em
     \fourteenpoint
  \noindent #1\par\endgroup }
\def\titlestyle#1{\par\begingroup \interlinepenalty=9999
     \leftskip=0.02\hsize plus 0.23\hsize minus 0.02\hsize
     \rightskip=\leftskip \parfillskip=0pt
     \hyphenpenalty=9000 \exhyphenpenalty=9000
     \tolerance=9999 \pretolerance=9000
     \spaceskip=0.333em \xspaceskip=0.5em
     \fourteenpoint
   \noindent #1\par\endgroup }
\def\spacecheck#1{\dimen@=\pagegoal\advance\dimen@ by -\pagetotal
   \ifdim\dimen@<#1 \ifdim\dimen@>0pt \vfil\break \fi\fi}
\def\section#1{\cleareqnumber \s@ctrue \global\advance\secnumber by1
   \par \ifnum\the\lastpenalty=30000\else
   \penalty-200\vskip\sectionskip \spacecheck\sectionminspace\fi
   \noindent {\caps\enspace\S\the\secnumber\quad #1}\par
   \nobreak\vskip\headskip \penalty 30000 }
\def\undertext#1{\vtop{\hbox{#1}\kern 1pt \hrule}}
\def\subsection#1{\par
   \ifnum\the\lastpenalty=30000\else \penalty-100\smallskip
   \spacecheck\sectionminspace\fi
   \noindent\undertext{#1}\enspace \vadjust{\penalty5000}}

\def\appendix#1{\cleareqnumber \s@cfalse \global\advance\appnumber by1
   \par \ifnum\the\lastpenalty=30000\else
   \penalty-200\vskip\sectionskip \spacecheck\sectionminspace\fi
   \noindent {\caps\enspace Appendix \char\the\appnumber\quad #1}\par
   \nobreak\vskip\headskip \penalty 30000 }
\def\ack{\par\penalty-100\medskip \spacecheck\sectionminspace
   \line{\fourteencp\hfil ACKNOWLEDGEMENTS\hfil}%
\nobreak\vskip\headskip }
\def\refs{\begingroup \par\penalty-100\medskip \spacecheck\sectionminspace
   \line{\fourteencp\hfil REFERENCES\hfil}%
\nobreak\vskip\headskip \frenchspacing }
\def\endrefs{\par\endgroup}
%
%
\newif\iffrontpage \frontpagefalse
\headline={\hfil}
\footline={\iffrontpage\hfil\else \hss\twelverm
-- \folio\ --\hss \fi }
%
%
\newskip\frontpageskip \frontpageskip=12pt plus .5fil minus 2pt
\def\titlepage{\global\frontpagetrue\hrule height\z@ \relax
               \pubblock\relax }
\def\endtitlepage{\vfil\break\clearfnotenumber\frontpagefalse}
\def\title#1{\vskip\frontpageskip\Titlestyle{\caps #1}\vskip3\headskip}
\def\author#1{\vskip.5\frontpageskip\titlestyle{\caps #1}\nobreak}
\def\and{\par\kern 5pt \centerline{\sl and}}
\def\andauthor{\vskip.5\frontpageskip\centerline{and}\author}

\def\address#1{\par\kern 5pt\titlestyle{\it #1}}
\def\andaddress{\par\kern 5pt \centerline{\sl and} \address}

\def\abstract#1{\par\dimen@=\prevdepth \hrule height\z@ \prevdepth=\dimen@
   \vskip\frontpageskip\spacecheck\sectionminspace
   \centerline{\fourteencp ABSTRACT}\vskip\headskip
   {\noindent #1}}

\def\email#1{\fnote{\tentt e-mail: #1\hfill}}
\def\newaddress#1{\fnote{\tenrm #1\hfill}}
%
%

%

%
\def\QMW{\address{%
   Department of Physics, Queen Mary and Westfield College\break
   Mile End Road, London E1 4NS, UK}}
%

%
%
\newcount\refnumber \def\clearrefnumber{\refnumber=0}  \clearrefnumber
\newwrite\R@fs                              
\immediate\openout\R@fs=\jobname.refs 
\def\closerefs{\immediate\closeout\R@fs} 
\def\refsout{\closerefs\refs
\unlockat
\input\jobname.refs
\lockat
\endrefs}
\def\refitem#1{\item{{\bf #1}}}
\def\ifundefined#1{\expandafter\ifx\csname#1\endcsname\relax}
\def\[#1]{\ifundefined{#1R@FNO}%
\global\advance\refnumber by1%
\expandafter\xdef\csname#1R@FNO\endcsname{[\the\refnumber]}%
\immediate\write\R@fs{\noexpand\refitem{\csname#1R@FNO\endcsname}%
\noexpand\csname#1R@F\endcsname}\fi{\bf \csname#1R@FNO\endcsname}}
\def\refdef[#1]#2{\expandafter\gdef\csname#1R@F\endcsname{{#2}}}
%
%
\newcount\eqnumber \def\cleareqnumber{\eqnumber=0}
\newif\ifal@gn \al@gnfalse  
\def\veqnalign#1{\al@gntrue \vbox{\eqalignno{#1}} \al@gnfalse}
\def\eqnalign#1{\al@gntrue \eqalignno{#1} \al@gnfalse}
\def\(#1){\relax%
\ifundefined{#1@Q}
 \global\advance\eqnumber by1
 \ifs@cd
  \ifs@c
   \expandafter\xdef\csname#1@Q\endcsname{{%
\noexpand\rm(\the\secnumber .\the\eqnumber)}}
  \else
   \expandafter\xdef\csname#1@Q\endcsname{{%
\noexpand\rm(\char\the\appnumber .\the\eqnumber)}}
  \fi
 \else
  \expandafter\xdef\csname#1@Q\endcsname{{\noexpand\rm(\the\eqnumber)}}
 \fi
 \ifal@gn
    & \csname#1@Q\endcsname
 \else
    \eqno \csname#1@Q\endcsname
 \fi
\else%
\csname#1@Q\endcsname\fi\global\let\@Q=\relax}
%
%
\newif\ifm@thstyle \m@thstylefalse
\def\mathstyle{\m@thstyletrue}
\def\proclaim#1#2\par{\smallbreak\begingroup
\advance\baselineskip by -0.25\baselineskip%
\advance\belowdisplayskip by -0.35\belowdisplayskip%
\advance\abovedisplayskip by -0.35\abovedisplayskip%
    \noindent{\caps#1.\enspace}{#2}\par\endgroup%
\smallbreak}
\def\m@kem@th<#1>#2#3{%
\ifm@thstyle \global\advance\eqnumber by1
 \ifs@cd
  \ifs@c
   \expandafter\xdef\csname#1\endcsname{{%
\noexpand #2\ \the\secnumber .\the\eqnumber}}
  \else
   \expandafter\xdef\csname#1\endcsname{{%
\noexpand #2\ \char\the\appnumber .\the\eqnumber}}
  \fi
 \else
  \expandafter\xdef\csname#1\endcsname{{\noexpand #2\ \the\eqnumber}}
 \fi
 \proclaim{\csname#1\endcsname}{#3}
\else
 \proclaim{#2}{#3}
\fi}
\def\Thm<#1>#2{\m@kem@th<#1M@TH>{Theorem}{\sl#2}}
\def\Prop<#1>#2{\m@kem@th<#1M@TH>{Proposition}{\sl#2}}
\def\Def<#1>#2{\m@kem@th<#1M@TH>{Definition}{\rm#2}}
\def\Lem<#1>#2{\m@kem@th<#1M@TH>{Lemma}{\sl#2}}
\def\Cor<#1>#2{\m@kem@th<#1M@TH>{Corollary}{\sl#2}}
\def\Conj<#1>#2{\m@kem@th<#1M@TH>{Conjecture}{\sl#2}}
\def\Rmk<#1>#2{\m@kem@th<#1M@TH>{Remark}{\rm#2}}
\def\Exm<#1>#2{\m@kem@th<#1M@TH>{Example}{\rm#2}}
\def\Qry<#1>#2{\m@kem@th<#1M@TH>{Query}{\it#2}}
%
%

%
\def\<#1>{\csname#1M@TH\endcsname}
%
%
\def\ref#1{{\bf [#1]}}
\def\ie{{\it i.e.\/}}
\def\nl{\hfil\break}
%
%

\def\lapprox{\hbox{\lower3pt\hbox{$\buildrel<\over\sim$}}}
\def\gapprox{\hbox{\lower3pt\hbox{$\buildrel<\over\sim$}}}
\def\quotient#1#2{#1/\lower0pt\hbox{${#2}$}}
\def\fr#1/#2{\mathord{\hbox{${#1}\over{#2}$}}}
\ifamsfonts
 \mathchardef\empty="0\hexmsb3F 
 \mathchardef\lsemidir="2\hexmsb6E 
 \mathchardef\rsemidir="2\hexmsb6F 
\else
 \let\empty=\emptyset
 \def\lsemidir{\mathbin{\hbox{\hskip2pt\vrule height 5.7pt depth -.3pt
    width .25pt\hskip-2pt$\times$}}}
 \def\rsemidir{\mathbin{\hbox{$\times$\hskip-2pt\vrule height 5.7pt
    depth -.3pt width .25pt\hskip2pt}}}
\fi
%
\def\to{\rightarrow}
%

%
%
\def\integ{\mathord{\bb Z}} 
%
%
\def\underrightarrow#1{\vtop{\ialign{##\crcr
      $\hfil\displaystyle{#1}\hfil$\crcr
      \noalign{\kern-\p@\nointerlineskip}
      \rightarrowfill\crcr}}} 
\def\underleftarrow#1{\vtop{\ialign{##\crcr
      $\hfil\displaystyle{#1}\hfil$\crcr
      \noalign{\kern-\p@\nointerlineskip}
      \leftarrowfill\crcr}}}  

\def\comm#1#2{\left[#1\, ,\,#2\right]}
%
\def\vder#1#2{{{{\delta}{#1}}\over{{\delta}{#2}}}}
\def\pder#1#2{{{\partial #1}\over{\partial #2}}}
\def\der#1#2{{{d #1}\over {d #2}}}
%
%

\def\NPB#1#2#3{{\sl Nucl. Phys.} {\bf B#1} (#2) #3}

\def\CMP#1#2#3{{\sl Comm. Math. Phys.} {\bf #1} (#2) #3}

\def\PLB#1#2#3{{\sl Phys. Lett.} {\bf #1B} (#2) #3}
\def\JMP#1#2#3{{\sl J. Math. Phys.} {\bf #1} (#2) #3}

\def\RMP#1#2#3{{\sl Rev. Mod. Phys.} {\bf #1} (#2) #3}

\def\PJAS#1#2#3{{\sl Proc. Jpn. Acad. Sci.} {\bf #1} (#2) #3}
\def\JPSJ#1#2#3{{\sl J. Phys. Soc. Jpn.} {\bf #1} (#2) #3}

\lockat
%

\let\pb=\anticomm

\def\spdo{{\hbox{S$\Psi$DO}}}

\let\Z=\integ
\def\sint{{\int_{B}}}

\def\cS{{\cal S}}
\def\fr#1/#2{\mathord{\hbox{${#1}\over{#2}$}}}
\def\half{\mathord{\fr1/2}}

\def\sres{{\rm sres\,}}

\def\Str{{\rm Str\,}}

\def\pair#1#2{\langle #1,#2\rangle} 

\def\ope#1#2{{{#2}\over{\ifnum#1=1 {x-y} \else {(x-y)^{#1}}\fi}}}

\def\W{{\ssf W}}

\def\kp{{\ssf KP}}

\def\SBKP{{\ssf SBKP}}
\let\bkp=\BKP
\def\SKP{{\ssf SKP}}\let\skp=\SKP

\def\KP{{\ssf KP}}\let\kp=\KP
\def\SKdV{{\ssf SKdV}}\let\skdv=\SKdV
\def\KdV{{\ssf KdV}}
\def\skp_2{{\ssf SKP_2}}

\refdef[MaRa]{Yu.~I.~Manin and A.~O.~Radul, \CMP{98}{1985}{65}.}
\refdef[DickeyIII]{L.A.~Dickey, {\sl Integrable equations and Hamiltonian
systems}, World Scientific.}
\refdef[KP]{E.~Date, M.~Jimbo, M.~Kashiwara, and T.~Miwa
\PJAS{57A}{1981}{387}; \JPSJ{50}{1981}{3866}.}
\refdef[DickeyKP]{L.~A.~Dickey, {\sl Annals of the New York Academy of
Science} {\bf 491} (1987) 131.}
\refdef[WKP]{J.M.~Figueroa-O'Farrill, J.~Mas, and E.~Ramos,
\PLB{266}{1991}{298}.}
\refdef[Yu]{F. Yu, {\sl Bi-Hamiltonian Structure of Super KP
Hierarchy}, UU-HEP-91/13,\nl ({\tt hep-th/9109009}).}
\refdef[N=1]{J.M.~Figueroa-O'Farrill and E.~Ramos,\CMP{145}{1992}{43}.}
\refdef[N=2]{J.M.~Figueroa-O'Farrill and E.~Ramos,\NPB{368}{1992}{361}.}
\refdef[Toda]{S. Komata, K. Mohri, and H. Nohara, \NPB{359}{1991}{168}.}
\refdef[virskdv]{P. Mathieu,\JMP{29}{1988}{2499}.}
\refdef[skpdos]{J.M. Figueroa-O'Farrill, J. Mas, and E. Ramos,
\RMP{3}{1991}{479}.}
\refdef[DargisMathieu]{P. Dargis and P. Mathieu, {\sl Nonlocal
conservation laws for supersymmetric KdV equations}, LAVAL-PHY -21/93,
({\tt hep-th/9301080}).}
\refdef[FiRa]{J.M. Figueroa-O'Farrill and E. Ramos,\PLB{262}{1991}{265}.}
\refdef[Oevel]{W. Oevel and Z. Popowicz,\CMP{139}{1991}{441}.}
\refdef[review]{P. Di Francesco, P. Ginsparg and J. Zinn-Justin,
{\sl 2-D gravity and random matrices}, ({\tt hep-th/9306153}).}
\overfullrule=0pt
\def\pubblock{ \line{\hfil\twelverm QMW-PH-94-3}
               \line{\hfil\twelverm February 1994}
               \line{\hfil hep-th/9402056}}
\titlepage
\title{On the Supersymmetric BKP-Hierarchy}
\author{Eduardo Ramos\email{e.ramos@qmw.ac.uk}}
\andauthor{Sonia Stanciu\email{s.stanciu@qmw.ac.uk}
\newaddress{On leave of absence from the Physikalisches Institut der
Universit\"at Bonn}}
\QMW
\abstract{ We prove that the supersymmetric BKP-hierarchy of Yu
($\SBKP_2$) is hamiltonian with respect to a nonlinear extension of
the $N=1$ Super-Virasoro algebra ($\W_{\SBKP}$) by fields of spin $k$,
where $k>3/2$ and $2k \equiv 0,3 \pmod 4$.  Moreover, we show how to
associate in a similar manner an $N=1$ $\W$-superalgebra with every
integrable hierarchy of the $\SKdV$-type.  We also show using dressing
transformations how to extend, in a way which is compatible with the
hamiltonian structure, the $\SBKP_2$-hierarchy by odd flows, as well
as the equivalence of this extended hierarchy to the $\SBKP$-hierarchy
of Manin-Radul.}
\endtitlepage
\section{Introduction}

One of the most fertile fields in theoretical physics in recent years
has been provided by the relationship between conformal field theory,
string theory, and integrable systems (for a review see for example
\[review]). But an extension of these results to the supersymmetric
arena has shown to be unexpectedly difficult.  In particular, the
connection between integrable hierarchies of the $\KdV$-type and
$W$-algebras, which is one of the pillars on which this field stands,
is not yet fully understood in the supersymmetric case.  It is on this
particular point that we have something to say.  But first, in order
to motivate our approach to this problem, it will be necessary to give
a brief historic description of supersymmetric hierarchies and
their relation to $\W$-superalgebras.

The first occurrence of $\W$-superalgebras in connection with
supersymmetric integrable systems appeared in the context of
supersymmetric Toda theories, where they play the role of integrals
of motion \[Toda]. It had also been known for a long time
\[virskdv] that the superVirasoro algebra provides a Poisson
structure for the supersymmetric KdV equation, but nevertheless a
general framework similar to the Adler-Gel'fand-Dickey (AGD) formalism
for generalized KdV hierarchies was still lacking.  In a series of
papers J.M.~Figueroa-O'Farrill and one of the authors developed
this formalism for the supersymmetric case. It was shown in
\[N=1] and \[N=2] how to construct $N=1$ and $N=2$ $\W$-superalgebras
associated with each {\it odd} super-Lax operator in
a $(1|1)$-superspace, but it was only when these superalgebras were
understood as Poisson structures that supersymmetry played a trick on
us. In contrast to the standard case, these $\W$-superalgebras were
not the Poisson structures associated with reductions of the
Manin-Radul $\SKP$-hierarchy \[MaRa].

It was realized in \[skpdos] how some of these problems could be
circumvented by the introduction of an {\it even} order super-Lax
operator.  The so called $\SKP_2$-hierarchy together with its
generalized KdV-type reductions are formally integrable and
bihamiltonian.  But it seems there is no free lunch in superspace, for
$\SKP_2$  the odd flows appear to be irremediably lost, and moreover
the associated Poisson structure does not display explicitly a
superconformal structure.  Nevertheless, it was shown in
\[skpdos]\[Oevel] that the $\SKdV$ equation appears as a reduction of
the $\SKP_2$ hierarchy and moreover that its natural Poisson
structure, {\ie} the superVirasoro algebra, appeared via hamiltonian
reduction from the Poisson structure of the $\SKP_2$ hierarchy.
Moreover, in a recent paper \[DargisMathieu] Dargis and Mathieu
showed how odd flows for the $\SKdV$ hierarchy could be defined in
terms of nonlocal odd conserved charges. These odd conserved
charges were generated by taking the fourth root of the associated
super-Lax operator, $L= D^4 + UD$. It seems from all of this that
things start to fall into place, but it is not clear from our previous
discussion that this is not more than a happy coincidence for a
particular reduction. As we will try to show in what follows, all of
these facts for the $\SKdV$ equation fall neatly in a general
structure that is supplied by the $\SBKP_2$ hierarchy introduced by Yu
in \[Yu].

Our original motivation for this paper stemmed from a very simple
remark that can be made by looking at this general picture, namely the
fact that the knowledge we have about each of the two supersymmetric
$\KP$ hierarchies, or equivalently even and odd Lax operators, seems
to cover the unknown side of the other. This suggested to us the main
theme of this paper: to establish a relation between $\SKP$ and
$\SKP_2$ and to combine their respective nice properties to make some
progress in our understanding of the $\SKP_2$ hierarchy.

More precisely the aim of this paper is twofold. On the one hand we
investigate the possibility of defining  consistent odd flows for the
$\SKP_2$ hierarchy.  In order to do this we shall briefly recall in
Section I the definition of the two supersymmetric $\KP$ hierarchies,
and we shall study under which conditions the Lax operator for $\SKP_2$ is
the square of the one for $\SKP$. This property is of the utmost
importance, because if fulfilled implies a one-to-one correspondence
between the Lax operators of both hierarchies, thus allowing us to induce
odd flows for $\SKP_2$ from the ones of $\SKP$, and establishing
the equivalence of both hierarchies.

On the other hand, we shall search for a reduction of $\SKP_2$
whose Poisson brackets are local and that exhibits explicitly a
superconformal structure. In section II we shall define a natural
reduction of the $\SKP_2$ hierarchy--the $\SBKP_2$ hierarchy--which
obeys all the desired properties.  We shall see that the associated
super-Lax operator is dressable, and therefore its even and odd flows
are equivalent to the ones of the $\SBKP$ hierarchy of  Manin and
Radul. In section III we shall display a local Poisson structure
explicitly exhibiting a superconformal structure, $\W_{\SBKP}$. This
is due to the fact that the Poisson structure of $\SBKP_2$ does not
come from a hamiltonian reduction of the one for $\SKP_2$, but rather
it is induced by a generalization to pseudodifferential operators of
the standard $N=1$ $\W$-superalgebras coming from odd super-Lax
operators! In particular it will become clear from this point of view
why the $\skdv$ equation was somehow misleading because of its
simplicity.  Finally, in section IV we shall consider once again the
$\skdv$ equation under the new light shed by our approach.  It will
then become obvious how to generalize the procedure in order to
construct $\skdv$-type reductions of the $\SBKP_2$ hierarchy
associated to finite dimensional $N=1$ superconformal algebras.
\section{Supersymmetric $\kp$ hierarchies. Dressability and odd flows}

Our main goal in this section is to define consistent odd flows for
the $\SKP_2$ hierarchy. The key ingredient in our approach is provided
by the supersymmetric analogue of the dressing transformations. We
will show that upon the condition of dressability the even flows
of $\SKP$ and
$\SKP_2$ hierarchies are equivalent. Moreover, it will become
obvious how to use that equivalence to induce odd flows for
$\SKP_2$ from those of $\SKP$.
We shall therefore start by defining the
$\SKP_2$ hierarchy, together with the $\SKP$ hierarchy, which will
play an essential role throughout this paper.

In the sequel we shall assume that the reader is familiar with the
basics of the supersymmetric formalism of pseudodifferential operators
($\spdo$'s). We shall only mention (for details see for instance
\[MaRa]) in order to fix the notation, that these $\spdo$'s are
defined on a $(1|1)$ superspace with coordinates  $(x,\theta)$. They
are given as formal Laurent series in $D^{-1}$ whose coefficients are
superfields, where $D^{-1}$ is a formal inverse to
$D=\partial_{\theta} +\theta\partial$. The multiplication for
$\spdo$'s is given by the following composition law (for any $n\in\Z$)
$$
D^n\Phi = \sum_{i=0}^\infty {n\brack {n-i}} (-1)^{|\Phi|(n-i)} \Phi^{[i]}
D^{n-i}\ ,\(compspdo)
$$
for $\Phi$ any superfield and where the superbinomial coefficients are given by
$$
{n\brack{n-i}} = \cases{0&for $i<0$ or $(n,i)\equiv (0,1)\pmod{2}$;\cr
{\left({{\left[\fr n/2\right]}\atop{\left[\fr n-i/2\right]}}\right)}&for
$i\geq 0$ and $(n,i)\not\equiv(0,1)\pmod{2}$.\cr}\(funnychoose)
$$

The $\SKP$ hierarchy is defined as the universal family of
isospectral flows deforming a $\spdo$ $\Lambda = D +\sum_{i\geq 1}V_i
D^{1-i}$. Its Lax flows are given by the following equations:
$$
\veqnalign{
D_{2i}\Lambda&= [\Lambda_-^{2i},\Lambda] =
                - [\Lambda_+^{2i},\Lambda]~,\(evenmr)\cr
D_{2i-1}\Lambda&= [\Lambda_-^{2i-1},\Lambda] =
                   -[\Lambda_+^{2i-1},\Lambda] +
                        2\Lambda^{2i}~.\(oddmr)\cr}
$$
In contrast to the nonsupersymmetric case this infinite family of
odd and even flows satisfy a nonabelian Lie superalgebra whose
commutation relations are
$$
[D_{2i},D_{2j}]=0~,\quad [D_{2i},D_{2j-1}]=0~,\quad
[D_{2i-1},D_{2j-1}]=2 D_{2i+2j-2}~.\(MRalgebra)
$$

The necessary and sufficient condition for $\Lambda$ to be dressable,
that is, for the existence of an even $\spdo$
$$
\phi =1+\sum_{i\geq 1}A_iD^{-i}~,\()
$$
in the superVolterra group, such that
$$
\Lambda=\phi D\phi^{-1}~.\()
$$
is $U_1^{[1]}+2U_2=0$. And it is easy to check that this condition is
indeed preserved by the flows.

On the other hand, the $\SKP_2$ hierarchy is defined as the universal
family of isospectral deformations of a $\spdo$ of the form
$$
L= D^2 + \sum_{i\geq 1}U_i D^{2-i}~,\()
$$
and its evolution is described by a commuting family of {\it even}
flows
$$
D_{2i} L = [L_-^i,L] =-[L_+^i,L]~.\(skp2flows)
$$
One can easily work out the dressability condition, which in this case
reads
$$
L=\phi\partial\phi^{-1},\(dressing)
$$
(where $\phi$ is again an element of the superVolterra group) and
obtain the necessary and sufficient condition $U_1=U_2=0$, which is
once again preserved by the flows.

As we have already mentioned in the introduction one of
the reasons for defining the $\SKP_2$
hierarchy was the fact that $L$ does not admit in general a (unique)
square root. Notice nevertheless that if we restrict ourselves to
dressable $L$'s then we can define
$$
L^{1/2} = \left(\phi D \phi^{-1}\right)~.\()
$$
In other words, the Lax operator of $\SKP_2$ admits a square root which
can be identified with the Lax operator of the $\SKP$ hierarchy. As it is
well known \[MaRa] the square root, if it exists, need not be unique.
Uniqueness can be nevertheless achieved in our case if we further
impose manifest supersymmetry as well as homogeneity with respect the
natural grading.  Let us recall that we can naturally asssociate a
degree to our Lax operators by declaring the degree of $D$ and $U_j$
to be $1/2$ and $1+j/2$ respectively. If we do so, the degree of $L$
is 1. If we require the degree of $L^{1\over 2}$ to be $1/2$ then the
square root is uniquely determined and equal to $\phi D\phi^{-1}$.
Moreover, if one works out explicitly the condition for the square
root of $L$ to exist, one obtains once more that $U_1$ and $U_2$ must
necessarilly vanish, and this is, as we have just shown, nothing but
the condition that $L$ be dressable.  Therefore, the square root of
$L$ exists if and only if it is dressable.

We are now only one step away of our odd $\SKP_2$ flows. Indeed, all
we need to is to show that every $\SKP$ flow induces a corresponding
flow in the $\SKP_2$, subject to the condition $L=\Lambda^2$. This can
be easily proven if we consider a generic flow of $\SKP$, which
looks like
$$
D\Lambda = [P,\Lambda]~,\()
$$
with $P$ a certain $\spdo$. Then we have
$$
\eqnalign{
D\Lambda^2 &= (D\Lambda)\Lambda + (-)^{|P|}\Lambda(D\Lambda)\cr
           &= [P,\Lambda^2]~,\()\cr}
$$
or in other words
$$
DL = [P,L]~.\(generic)
$$
Let us show that the converse is also true, although it requires a
little more algebra to prove it. Consider a generic $\SKP_2$ flow, of
the form \(generic) and take into account the fact that $L=\Lambda^2$.
Then we get
$$
(D\Lambda - [P,\Lambda])\Lambda + (-)^{|P|}\Lambda(D\Lambda -
[P,\Lambda]) = 0~.\()
$$
If we assume for a moment that the expression in the paranthesis does
not vanish but rather $D\Lambda - [P,\Lambda]=B_N D^N
+B_{N-1}D^{N-1}+\ldots$ , for some $N\in\Z$, we obtain that the
leading coefficient $B_N=0$. Hence
$$
D\Lambda = [P,\Lambda]~.\()
$$

{}From this it is obvious that, upon dressability, the even flows of
$\SKP_2$ \(skp2flows) are in one-to-one correspondence with the even
flows of $\SKP$ \(evenmr). Moreover, we can now use the odd flows
\(oddmr) of $\SKP$ to induce the corresponding odd flows on $\SKP_2$,
{\ie},
$$
D_{2i-1}L= [L_-^{i-\half },L]~.\(induced)
$$

It is important to remark that although the odd flows are explicitly
local in terms of the $\SKP$ variables, this is not so (as a
straightforward computation reveals) when they are
written in terms of the $\SKP_2$ variables. This was already pointed
out by Dargis and Mathieu for the $\skdv$ case. One could therefore
ask whether there is no other way of providing the $\SKP_2$ hierarchy
with explicitly {\it local} odd flows. But a brute force computation
for the first few flows shows that this is indeed impossible.
\section{The $\SBKP_2$ Hierarchy}

It is well known since the work in \[skpdos]\[Oevel]\[Yu]
that the $\SKP_2$ hierarchy admits a bihamiltonian structure. It is
then natural to ask which is the hamiltonian structure induced via
hamiltonian reduction on the space of dressable operators.
Unfortunately, a straightforward computation using Dirac brackets
shows that the induced Poisson algebra is nonlocal. But, as already
mentioned in the introduction, there exists at least a particular
reduction of the $\SKP_2$ with has all the desired properties; {\ie},
the $\skdv$ hierarchy. In what follows, we will use this example as a
guiding principle in our search of a ``nice'' reduction of the
$\SKP_2$ hierarchy.

Let us recall that the $\SKdV$ hierarchy is obtained
from the Lax equations associated with the operator $L=D^4 +UD$.
This Lax operator can be described in a more invariant way as the
unique superdifferential operator of order four satisfying the
constraint
$$
L^* = D L D^{-1}~,\(constraintskdv)
$$
where $*$ is the unique involution in the ring of \spdo's satisfying
the basic properties
\item{($a$)}{$D^* = -D$,}
\item{($b$)}{$f^* = f$, for any differential polynomial $f$; and,}
\item{($c$)}{$(PQ)^* = (-1)^{|P||Q|} Q^*P^*$, for all homogeneous
$P,Q\in\cS$.}

Moreover, the involution $*$ enjoys the following additional properties:
\item{(1)}{If $P\in\cS$ is homogeneous and invertible, $(P^{-1})^* = (-1)^{|P|}
(P^*)^{-1}$.}
\item{(2)}{For all $p\in\integ$, $(D^p)^* = (-1)^{p(p+1)\over 2} D^p$.}
\item{(3)}{For all $P\in\cS$, $(P_{\pm})^* = (P^*)_{\pm}.$}
\item{(4)}{For all $P\in\cS$, $\sres\,P^* = \sres\,P$ (in particular,
$\Str\,P^* = \Str \,P$).

The condition \(constraintskdv) is nothing but the supersymmetric
analogue of the constraint used in \[KP] to define the $\bkp$
hierarchy\fnote{It is amusing that from the Grassmannian approach to
the $\KP$ hierarchy this reduction amounts to going from Dirac to
Majorana fermions.}, it is therefore natural to impose such constraint
at the level of the $\SKP_2$ Lax operator; {\ie},
$$
L^* = -D L D^{-1}~.\(constskp)
$$
In fact, this reduction ($\SBKP_2$)
was shown in \[Yu] to be consistent with the Lax flows as long
as only half of them are considered. Let us for the sake of
completeness recall the proof.
It is clear from \(skp2flows)
that the consistency of this reduction for the
$i$-th flow boils down to the
condition that
$$- D \comm{L}{L^i_-} D^{-1} = \comm{L}{L^i_-}^* ,\()$$
which is only satisfied provided
$$ (D L^i D^{-1})_- = -(-)^i D L^i_- D^{-1}.\(zeromode)$$
On the one hand, the leading term of the above equation clearly implies that
$i$ should be an odd integer. On the other hand,
it is a simple computational matter to check that for an arbitrary
{\spdo}  $A=\sum_j A_j D^j$ the relation $(D A D^{-1})_- = D A_- D^{-1}$
holds if and only if $A_0 =0$, in other words if
$\sres L^i D^{-1} =0$.
But this is automatically
fulfilled for $i$ an odd integer. Indeed
$$\eqnalign{\sres L^i D^{-1} &=\sres (L^i D^{-1})^*\cr
&=\sres D^{-1} (L^*)^i\cr
&=(-)^i \sres L^i D^{-1}.\(zeroresidue)\cr}$$

In order to understand better which is the meaning of the constraint
at the level of the $U_j$'s it will show convenient to write the
operator $L$ in a different basis. Without loss of generality $L$
can be written as
$$L= D^2 + {1\over 2}\sum_{k\geq 0}
\pb{U_k}{D^{-k}}D,\(newbasis)$$
where $\pb{\ }{\ }$ stands for the graded anticommutator, that
is $\pb{A}{B} = AB + (-)^{|A||B|} BA$.
In term of these variables
$$L^* = -D^2 + {1\over 2}\sum_{k\geq 0} (-)^{k(k-1)\over 2}
D\pb{U_k}{D^{-k}}.\()$$
Therefore we can read directly from this that consistency
with the constraint
\(constskp) requires the index $k$ to
take values in the set ${\cal I}_k$, where ${\cal I}_k
=\{ k | k\in\integ_+ \  {\rm and}\ k\equiv 2,3 \pmod 4 \}$.

It is sufficient to write down the first few terms of $L$ to realize
that this reduction sets the first two fields equal to zero.
Hence $L$ does fulfill the requisites to be a dressable operator, and
we can apply the machinery developed in the previous section to define
the odd flows associated with the half-integer powers of the Lax
operator.

The odd flows for the $\SBKP_2$ will be induced from the ones of
$\SKP$ subject to the condition that $\Lambda^*=-D\Lambda D^{-1}$
($\SBKP$). In order to see which odd flows of $\SKP$ preserve this
constraint we can proceed as before.  The consistent odd flows are
then given by
$$
D_{4i-1} \Lambda = [\Lambda_-^{4i-1},\Lambda].\()
$$
Hence the full-fledged $\SBKP_2$ flows read
$$
\eqnalign{
D_{4i-2} L =& [L_-^{2i-1},L],\(even)\cr
D_{4i-1} L =& [L_-^{2i-\half},L].\(odd)\cr}
$$

Moreover, from their equivalence to the $\SBKP$ flows it follows that
they obey the following Lie superalgebra:
$$
\comm{D_{4i-2}}{D_{4j-2}}=0,\quad\comm{D_{4i-2}}{D_{4j-1}}=0,
\quad\comm{D_{4i-1}}{D_{4j-1}} = 2 D_{4i+4j-2}~.\()
$$

\section{Hamiltonian structure for $\SBKP_2$ and $\W_{\SBKP}$}

One would naively expect that the hamiltonian structure for the
$\SBKP_2$-hierarchy is simply provided by a suitable reduction of the
hamiltonian structure of the $\SKP_2$-hierarchy, but a simple computation
shows that things become more complicated by the fact that the
constraints seem to be formally first-class. Although in standard
finite dimensional examples there is a well developed machinery to
deal with such constraints, in this particular case the existence of
an infinite number of them seriously complicates a similar analysis.
At this point one could think that there is no simple way to obtain a
hamiltonian structure for $\SBKP$ from the standard AGD approach. But
a simple observation will come to our rescue.

{}From the work in \[N=1] is known that there is a natural $N=1$
$\W$-superalgebra associated with each differential superLax operator
$\Lambda$ of order $2k+1$ subject to the constraint $\Lambda^* =
-(-)^k\Lambda$.
In complete analogy with the results for the bosonic case
\[DickeyKP]\[WKP] this
hamiltonian structure can be extended to {\spdo}'s of order
$2k+1$ subject to the same constraint.
We will show that the Poisson algebra thus obtained ,$\W_{\SBKP}$,
will induce, after a certain geometric procedure, a Poisson structure
for $\SBKP_2$. But before doing so it will be necessary to recall the
standard formalism.

\subsection{Supersymmetric Adler-Gel'fand-Dickey formalism for \spdo's}}

It is our immediate purpose to define Poisson brackets in the space of
(unconstrained) \spdo's of the form
$$\Lambda = D^{n} +\sum_{j=1}^{\infty} V_j D^{n-j}.\(onesuperw)$$
These spaces are infinite-dimensional
manifolds and their rigorous geometric treatment is beyond our scope.
Fortunately these spaces can be endowed with a ``formal'' geometry which is
sufficient for the rigorous treatment of Poisson brackets and
their flows \[DickeyIII].  As usual, this formal geometry consists in
the algebraization of the necessary geometric notions.  Therefore we
proceed to define algebraically the ingredients needed to define
Poisson structures: functions, vector fields, and 1-forms.

We will define Poisson brackets on functions of the form:
$$F[\Lambda] = \sint f(V)\ ,\()$$
where $f(V)$ is a homogeneous (under the $\integ_2$ grading) differential
polynomial of the $V$'s and $\sint$ is defined as follows: if $U_i=u_i+\theta
v_i$, and $f(U)= a(u,v) + \theta b(u,v)$, then $\sint f(U) = \int b(u,v)$,
where the precise meaning of integration will depend on the context,
but it can be regarded more abstractly as a linear map annihilating
derivatives so that we can ``integrate by parts".

Vector fields are parametrized by infinitesimal deformations
$\Lambda\mapsto \Lambda +
\epsilon A$ where $A=\sum A_l D^l$ is a homogeneous differential operator of
order at most $n$. We denote the space of such operators by $S_n$.  We
do not
demand that $A$ have the same parity as $\Lambda$ since we can have
both  odd and
even flows.  To such an operator $A\in S_n$ we associate a vector field $D_A$
as follows.  If $F=\sint f$ is a function then
$$\veqnalign{D_A\,F &\equiv \left.\der{}{\epsilon} F[\Lambda+\epsilon
A]\right\vert_{\epsilon=0}\cr
&=(-1)^{|A|+n}\sint\sum_{k=1}^{\infty}\sum_{i=0}^\infty
(-1)^{(|A|+n)i}A_k^{[i]}\pder{f}{V_k^{[i]}}\ ,\()\cr}$$
with $V_k^{[i]}=D^iV_k$ and the same for $A_k^{[i]}$.
Integrating by parts we can write this as
$$D_A\,F = (-1)^{|A|+n}\sint \sum_{k=1}^{\infty} A_k \vder{f}{V_k}\
,\(nose)$$
where the Euler variational derivative is given by
$$\vder{\phantom{f}}{V_k}= \sum_{i=0}^\infty (-1)^{|V_k|i + i(i+1)/2} D^i
\pder{\phantom{f}}{V_k^{[i]}}\ .\()$$

Since vector fields are parametrized by $S_n$, it is natural to think
of 1-forms as pa\-ram\-e\-trized by the dual space to $S_n$, with the dual
pairing given by the Adler supertrace.
Let us then define 1-forms as the space $S_n^*$ of \spdo's of the
form $X=\sum^{\infty}_{k=1} D^{k-n-1} X_k$, whose pairing with a
vector field $D_A$, with $A=\sum A_kD^k$, is given by
$$\pair{D_A}{X}\equiv (-1)^{|A|+|X|+n+1}\Str (AX) = (-1)^{|A|+n}\sint
\sum_{k=1}^{\infty}(-1)^k A_k\,X_k\ ,\()$$
which is nondegenerate.  The choice of signs has been made to avoid
undesirable signs later on.  Given a function $F=\sint f$ we define its
gradient $dF$ by $\pair{D_A}{dF}=D_AF$ whence, comparing with \(nose),
yields
$$dF = \sum_{k=1}^{\infty} (-1)^k D^{-k-1} \vder{f}{V_k}\ .\()$$

It is familiar from classical mechanics that to every function $f$ one can
associate a hamiltonian vector field $\xi_f$ in such a way that $\xi_f\,g
=\pb{f}{g}$.  The hamiltonian vector field $\xi_f$ is obtained from $df$ by
a tensor $\Omega$ mapping 1-forms to vector fields, so that the
Poisson bracket of two functions is given by $\pb{f}{g}=\pair{X_f}{dg} =
\pair{\Omega(df)}{dg}$, with $\pair{}{}$ being the natural pairing between
vector fields and 1-forms.  In local coordinates, $\Omega$ coincides with the
fundamental Poisson brackets.  In other words, the map $\Omega$ carries the
same information as the Poisson brackets.

In analogy, we define Poisson brackets on the space of supersymmetric Lax
operators by defining a map $J:S_n^* \to S_n$ in such a way that the Poisson
bracket of two functions $F$ and $G$ is given by
$$\pb{F}{G} = D_{J(dF)}G = \pair{D_{J(dF)}}{dG} = (-1)^{|J|+|F|+|G|+n+1}\,
\Str(J(dF)dG)\ .\(AdlerPoisson)$$
The map $\Omega$ in this case is $dF\mapsto D_{J(dF)}$; although, because of
the rather formal nature of our geometrical setting, it is the map $J$ that
plays the more relevant r\^ole.  Demanding that the Poisson brackets defined
by $J$ obeys the Jacobi identity imposes strong restrictions on the
allowed maps $J$.  Maps obeying these conditions are often called
``hamiltonian''.

The proof that the map $J$ given by
$$J(X)=(\Lambda X)_+\Lambda - \Lambda(X\Lambda )_+ =
\Lambda(X\Lambda )_- -(\Lambda X)_-\(Adlermap)$$
is hamiltonian follows mutatis mutandi the one of \[N=1], so it will
not be repeated here.

\subsection{$\W_{\SBKP}$}

Let $M_k$ denote the space of Lax operators of the form
$\Lambda=D^{2k+1}+\cdots$, and
let $\tilde M_k$ denote the submanifold of symmetric operators
$\Lambda^*=-(-1)^k \Lambda$.  As we will
see this submanifold inherits a well-defined
Poisson bracket from that in $M_k$.  To describe the bracket we first need to
identify the vector fields and the 1-forms on $\tilde M_k$ as subobjects of the
corresponding objects in $M_k$.
The vector fields of $\tilde M_k$ will be parametrized
by the deformations of $\Lambda$ that remain in $\tilde M_k$.
That is, deformations of a
symmetric Lax operator $\Lambda$ which keep it symmetric.
These are clearly the
{\spdo}'s of order at most $2k$ obeying the same symmetry property
of $\Lambda$.  As explained for example in \[N=2], 1-forms on
$\tilde M_k$ must be
chosen to be those 1-forms on $M_k$ which are mapped (via the
hamiltonian map $J$) to
vector fields tangent to $\tilde M_k$.
In other words, 1-forms on $\tilde M_k$ are \spdo's
$X=\sum_l D^{l-2k-2}X_l$ satisfying that $J(X)^* = -(-1)^k J(X)$.
Computing this we find
$$\veqnalign{J(X)^* =& \left[ (\Lambda X)_+
\Lambda - \Lambda(X\Lambda)_+\right]^*\cr
  =& (-1)^{|X|+1} \left[\Lambda^*(\Lambda X)_+^* -
(X\Lambda)_+^* \Lambda^*\right]\cr
  =& - \left[ \Lambda^* (X^*\Lambda^*)_+ -
(\Lambda^*X^*)_+\Lambda^*\right]\cr
  =& (\Lambda X^*)_+ \Lambda - \Lambda (X^*\Lambda)_+\cr
  =& J(X^*)\ ,\cr}$$
whence $X$ must have the same symmetry properties of
$\Lambda$, namely $X^* = -(-1)^k
X$, for it to be a 1-form on $\tilde M_k$.
It is easy to verify that these 1-forms
are non-degenerately paired with the vector fields tangent to
$\tilde M_k$.  In fact,
since $\Str\,(AX) = \Str\,(A^*X^*)$ we see that the supertrace pairs
up 1-forms and vector fields of the same symmetry properties.
Therefore the Poisson
bracket of two functions $F=\sint f$ and $G=\sint g$ on
$\tilde M_k$ is obtained from \(AdlerPoisson)
(with $J$ given by \(Adlermap)) by simply requiring that $dF$ and
$dG$ have the correct symmetry properties: $(dF)^*=-(-1)^k dF$ and the same for
$dG$.

It is now simple to check that the induced fundamental Poisson brackets on
$\tilde M_k$ contain
the $N=1$ supervirasoro algebra as a subalgebra.
Notice that because of particular form of the constraint the Lax
operator $\Lambda$ takes always the form
$$\Lambda = D^{2k+1} + V_3  D^{2k-2} +\cdots \ ,\()$$
where $V_3 $ is a field of weight $3/2$. A direct computation of the
Poisson bracket of $V_3 $ with itself yields
$$\pb{V_3 (Z)}{V_3 (Z')}\!=\!\left( {k(k+1)\over 4} D^5 +{3\over 2}
V_3 (Z) D^2 +{1\over 2} V_3 '(Z) D +\! V_3 ''(Z)\!\right) \delta(Z-Z'),$$
where the operator is acting at point $Z$, and $\delta(Z-Z')$ stands
for the standard delta function in superspace.

\subsection{Back to $\SBKP_2$}

Let us define now the map $\varphi$ from the space of $\SKP$
operators,
$M_0$, to the space of $\SKP_2$ operators, $N$, given by
$$\varphi (\Lambda) = \Lambda D=L.\(map)$$
Now we can compute the image under $\varphi$ of the
subspace $\tilde M_0$ defined by the condition\fnote{
The reader should not confuse this $\Lambda$ with the Lax operator
for $\SBKP$ which obeys a different constraint.} $\Lambda^* =-\Lambda$.
If we denote $\bar N =\varphi (\tilde M_0)$, then it is clear from
\(map) that if $L\in\bar N$ then $L^* =-DLD^{-1}$ which is precisely
the constraint imposed in $\SBKP_2$.  From the geometric formalism
developed above we know that the Adler map can be understood as
tensorial map from 1-forms to vector fields. Therefore, we can use the
map $\varphi$ to induce an Adler-type map on the space of $\SBKP_2$
operators from the standard Adler map $J$ in $\tilde M_0$
as follows. If $X$ is any 1-form in $\bar N$, {\ie}
$$X^* =(-)^{|X|} DXD^{-1},\()$$
we first pull it back to $\tilde M_0$ via $\varphi^*$ to obtain
$$Y=\varphi^*(X) = DX.\()$$
Notice that $Y^*=-Y$ and therefore is a 1-form on $\tilde M_0$. We
can then apply the pushforward $\varphi_*$ to the vector field that
is obtained by the application of the Adler map $J$ to $Y$.
Therefore the induced Poisson brackets will be the ones associated
to the Adler- type map $J_{\SBKP_2}$ defined by
$$J_{\SBKP_2}(X) = \left( \varphi_* \circ J \circ \varphi^*
\right).\(newmap)$$
A straightforward computation now yields
$$J_{\SBKP_2} (X) = (LX)_+L - LD^{-1}(DXLD^{-1})_+ D.\(newmapdos)$$
And \(AdlerPoisson) with $J=J_{\SBKP_2}$ define consistent Poisson
brackets among two functions $F=\sint f$ and $G=\sint g$ on $\bar N$
by simply requiring that their differential have the correct symmetry
properties; {\ie}, $(dF)^* = (-)^{|dF|} DdFD^{-1}$, and equivalently
for $dG$. Notice that despite its appearance \(newmapdos) induce
perfectly local Poisson brackets on $\bar N$. This can be most
easily seen by realizing that at the component level the map $\varphi$ is
simply given by $V_j\mapsto U_j$, therefore the locality of the
fundamental Poisson brackets on $\tilde M_0$ ensures the locality of
the fundamental Poisson brackets among the $U_j$.

In order to show that $J_{\SBKP_2}$ provides a hamiltonian structure
for the $\SBKP_2$ flows let us first consider the following set
of hamiltonian functions
$$H_{4k -2} = {1\over {2k-1}}\sint L^{2k-1}.\(evenham)$$
Their gradients are given by
$$dH_{4k-2}=L^{2k-2},\(gradevenham)$$
and they obey $(dH_{4k-2})^* = DdH_{4k-2}D^{-1}$, so they are well
defined 1-forms on $\bar N$. From \(newmapdos) and \(zeroresidue)
follows that
$$D_{4k-2}L=J_{\SBKP_2}(dH_{4k-2}),\(evenfinalshit)$$
as required. To prove that the odd flows are hamiltonian with
respect to $J_{\SBKP_2}$ is a much more delicate matter.
Notice that the natural odd hamiltonians are provided by
$$H_{4k-1}={1\over {2k-{1\over 2}}}\sint L^{2k-{1\over
2}},\(oddham)$$
and their integrands are nonlocal functionals of the $U_j$, therefore
the formalism developed for differential polynomials of the $U_j$
should be extended to integro-differential polynomials. Since this is
point is very technical in nature, we relegate the details to a
different forum, and for the time being we will proceed formally, but
correctly, to obtain
$$dH_{4k-1} = L^{2k-{3/2}},\()$$
which defines an odd 1-form in $\bar N$. Finally
$$D_{4k-1}L = J_{\SBKP_2}(dH_{4k-1}),\(oddfinalshit)$$
showing that the odd flows are also hamiltonian with respect
$\W_{\SBKP}$.

\section{$\SKdV$ revisited and more}

The purpose of this section is to show how the formalism developed
for $\SBKP_2$ can be equally applied to its reductions of the
$\SKdV$-type.

Let us consider the following constraints on the $\SBKP_2$ operator
$L$
$$(L^n)_-=0,\quad {\rm and}\quad \sres L^n D^{-1}=0,\(othercons)$$
for $n$ is a positive integer. Notice that the second constraint
is forced on us for consistency with $(L^n)^* = (-)^n DL^nD^{-1}$.

{}From the fact that the $\SBKP_2$ flows act as a superderivation it
follows that
$$\eqnalign{D_{4k-2} L^n =&
\comm{L^n}{L^{2k-1}_+},\(equnorep)\cr
D_{4k-1} L^n =& \comm{L^n}{(L^{2k-{1\over 2}})_+},\(equdosrep)\cr
}$$
making obvious that the first constraint is preserved by
the flows. The compatibility of the second constraint follows from
the fact that $L^{2k-1}_+$ as well as
$L^{2k -{1\over 2}}_+$ are differential operators with no free term.

The constraints \(othercons) can be suggestively written as
$L=\varphi (\Lambda),$
where $\varphi$ was defined in \(map) and $\Lambda$ is a superdifferential
symmetric operator of order $2n-1$; {\ie}, $\Lambda^* = (-)^n\Lambda$.
{}From the construction of the previous section it follows that we can
provide these subhierarchies with a hamiltonian structure induced via
$\varphi$ from the one in the space of $\Lambda$'s \[N=1].
Explicitly, if we denote by $L_{(n)}$ the $n$-th power of the
$\SBKP_2$ operator L subject to the constraints \(othercons), we
obtain, following the same procedure as before, that
$$J_{(n)} (X) = (L_{(n)}X)_+L_{(n)} - L_{(n)}D^{-1}(DXL_{(n)}
D^{-1})_+ D,\(strucskdv)$$
where $X$ is a 1-form in the space of operators $L_{(n)}$; that is,
$X^*=(-)^{|X|+n+1}DXD^{-1}$.

In the particular case when $n=2$, $L_{(2)}$ is nothing but the $\SKdV$
operator $D^4 + UD$, and the diligent reader can check that our formalism
provides a hamiltonian structure for the $\SKdV$ equation and its
``Dargis-Mathieu'' extension by odd flows. The Poisson algebra is
nothing but the $N=1$ superVirasoro algebra with central charge ${1\over
2}$ associated to the Lax operator $D^3 + U$. Notice that in contrast
with the formalism in \[skpdos] the Poisson structure is obviously local
and can be easily generalized to the $\SKdV$-type reductions
associated with higher values of $n$.
\vskip 1truecm

\ack

We would like to thank J.M.~Figueroa-O'Farrill and J.~Mas for many
useful and enlightening conversations on the subject.

\refsout
\bye